\documentclass{PoS}
\usepackage{graphicx}
\usepackage[utf8]{inputenc}
\usepackage{here}
\usepackage{color}
\usepackage{cite}

\usepackage{tikz}
\usetikzlibrary{positioning,arrows}
\usetikzlibrary{decorations.pathmorphing}
\usetikzlibrary{decorations.markings}
\usepackage{pgfplots}
\usepackage{xparse}
\usetikzlibrary{positioning,arrows,patterns}
\usetikzlibrary{decorations.markings}
\usetikzlibrary{calc}

\def\bc{\begin{center}}
\def\ec{\end{center}}
\def\be{\begin{equation}}
\def\ee{\end{equation}}
\def\beqn{\begin{eqnarray}}
\def\eeqn{\end{eqnarray}}
\def\no{\nonumber}
\def\nn{\no\\}
\def\eqn#1{(\ref{#1})}
\def\ba{\begin{array}{c}}
\def\ea{\end{array}}
\def\bat{\begin{array}{cc}}
\def\bi{\begin{itemize}}
\def\ei{\end{itemize}}
\def\cA{{\cal A}}
\def\cL{{\cal L}}

\def\cO{{\cal O}}

\title{Current status of $\varepsilon'/\varepsilon$ in the Standard Model }

\ShortTitle{Current status of $\varepsilon'/\varepsilon$ in the Standard Model}

\author{\speaker{Héctor Gisbert}\\
        Departament de F\'\i sica Te\`orica, IFIC, Universitat de Val\`encia -- CSIC\\
Apt. Correus 22085, E-46071 Val\`encia, Spain\\
        E-mail: \email{hector.gisbert@ific.uv.es}}

\abstract{A recent lattice prediction of $\varepsilon'/\varepsilon$, with a 2.1 $\sigma$ deviation from the experimental value, has triggered several studies of possible contributions from physics beyond the Standard Model. We have recently updated the SM prediction, including all known short-distance and long-distance contributions, our result $\mbox{Re}\left(\varepsilon'/\varepsilon\right) = (15 \pm 7)\cdot 10^{-4}$ \cite{Gisbert:2017vvj} is in complete agreement with the experimental measurement. In addition, we emphasize on the importance of the long-distance re-scattering of the final pions in $K \to \pi \pi$ for a correct prediction of $\varepsilon'/\varepsilon$.}

\FullConference{XIII Quark Confinement and the Hadron Spectrum - Confinement2018\\
		31 July - 6 August 2018\\
		Maynooth University, Ireland}

\begin{document}

\section{Introduction}

Since long time ago, the theoretical and experimental sides have been motivated to study  the CP violating ratio $\varepsilon'/\varepsilon$, because it constitutes a fundamental test for our understanding of flavour-changing phenomena.
The present experimental world average \cite{Batley:2002gn,Lai:2001ki,Fanti:1999nm,Barr:1993rx,Burkhardt:1988yh,Abouzaid:2010ny,AlaviHarati:2002ye,AlaviHarati:1999xp,Gibbons:1993zq},
\be\label{eq:exp}
\mathrm{Re} \left(\varepsilon'/\varepsilon\right)\; =\;
(16.6 \pm 2.3) \cdot  10^{-4}\, ,
\ee
demonstrates the existence of direct CP violation in the decay transitions $K^0\to 2\pi$.

The theoretical prediction of $\varepsilon'/\varepsilon$ has been the subject of many debates because the first next-to-leading order (NLO) calculations \cite{Buras:1993dy,Buras:1996dq,Bosch:1999wr,Buras:2000qz,Ciuchini:1995cd,Ciuchini:1992tj} obtained Standard Model (SM) values one order of magnitude smaller than \eqn{eq:exp}. However, it was soon realized that the final-state interactions (FSI) in $K\to\pi\pi$ are key for a reliable prediction\cite{Pallante:1999qf,Pallante:2000hk}. Once all relevant contributions were taken into account, the theoretical prediction was found to be in good agreement with the experimental value although with a large uncertainty of non-perturbative origin \cite{Pallante:2001he}.

Recently, the RBC-UKQCD lattice collaboration has published a prediction for the direct CP violation ratio, $\mathrm{Re}(\varepsilon'/\varepsilon) = (1.4 \pm 6.9) \cdot 10^{-4}$ \cite{Bai:2015nea,Blum:2015ywa}, which is in clear conflict with the experimental value in  Eq. \eqn{eq:exp}. This result has triggered many new studies of possible contributions from physics beyond the SM in order to explain this ``anomaly''.\footnote{Several studies can be found at the Inspire data basis. We refrain to quote them here} However, the same lattice simulation fails in its attempt to reproduce the $(\pi\pi)_I$ phase shifts, which provide a quantitative test of the lattice results. Although, the extracted $\delta_2$ is only $1\,\sigma$ away from its physical value, the lattice analysis of Ref.~\cite{Bai:2015nea} finds a result for
$\delta_{0}$ which disagrees with the experimental value by $2.9\,\sigma$, a much larger discrepancy than the one quoted for $\varepsilon'/\varepsilon$. Therefore, this discrepancy cannot be taken as evidence of new physics. Lattice practitioners are working on a better lattice understanding of the pion dynamics and improved results are expected soon \cite{Kelly:CKM2018}.

Since the publication of the SM $\varepsilon'/\varepsilon$ prediction in Ref.~\cite{Pallante:2001he}, there have been a lot of improvements in the isospin-breaking corrections \cite{Cirigliano:2003nn,Cirigliano:2003gt,Cirigliano:2009rr},  the quark masses \cite{Aoki:2016frl} and a better understanding of the chiral perturbation theory Low-Energy-Constants (LECs) \cite{Ecker:1988te,Ecker:1989yg,Pich:2002xy,Cirigliano:2006hb,Kaiser:2007zz,Cirigliano:2004ue,Cirigliano:2005xn,RuizFemenia:2003hm,Jamin:2004re,Rosell:2004mn,Rosell:2006dt,Pich:2008jm,GonzalezAlonso:2008rf,Pich:2010sm,Bijnens:2014lea,Rodriguez-Sanchez:2016jvw,Ananthanarayan:2017qmx}. Therefore, the current situation makes mandatory to revise and update the analytical SM calculation of $\varepsilon'/\varepsilon$ \cite{Pallante:2001he}. In the following, we give a brief summary of the new determination of $\varepsilon'/\varepsilon$ from Ref.~\cite{Gisbert:2017vvj}.

\section{Direct CP violation ratio $\boldmath \varepsilon'/\varepsilon$ }
\label{sec:anatomy}

The kaon decay amplitudes can be decomposed as
\begin{eqnarray}  
A(K^0 \to \pi^+ \pi^-) &=&  
\cA_{1/2} + {1 \over \sqrt{2}} \left( \cA_{3/2} + \cA_{5/2} \right) 
\; =\; 
 A_{0}\,  e^{i \chi_0}  + { 1 \over \sqrt{2}}\,    A_{2}\,  e^{i\chi_2 } \, ,
\no\\[2pt]
A(K^0 \to \pi^0 \pi^0) &=& 
\cA_{1/2} - \sqrt{2} \left( \cA_{3/2} + \cA_{5/2}  \right) 
\; =\;
A_{0}\,  e^{i \chi_0}  - \sqrt{2}\,    A_{2}\,  e^{i\chi_2 }\, ,
\no\\[2pt]
A(K^+ \to \pi^+ \pi^0) &=&
{3 \over 2}  \left( \cA_{3/2} - {2 \over 3} \cA_{5/2} \right) 
\; =\;
{3 \over 2}\, A_{2}^{+}\,   e^{i\chi_2^{+}},
\label{eq:2pipar}
\end{eqnarray}
where the complex amplitudes $\cA_{\Delta I}$ are generated by the $\Delta I = \frac{1}{2},
\frac{3}{2},\frac{5}{2}$ components of the electroweak effective Hamiltonian, in the limit of isospin conservation. The $A_0$, $A_2$ and $A_2^+$ amplitudes are real and positive in the CP-conserving limit. Moreover, in the isospin limit, $A_0$ and $A_2=A_2^+$ denote the decay amplitudes into $(\pi\pi)_{I=0,2}$ states and the phase differences $\chi_0$ and $\chi_2=\chi_2^+$ are the corresponding S-wave scattering phase shifts. From the measured $K\to\pi\pi$ branching ratios, one gets \cite{Antonelli:2010yf}:
\begin{eqnarray}
A_0 &=& (2.704 \pm 0.001) \cdot 10^{-7} \mbox{ GeV}, \nn
A_2 &=& (1.210 \pm 0.002) \cdot 10^{-8} \mbox{ GeV}, \nn
\chi_0 - \chi_2 &=& (47.5 \pm 0.9)^{\circ}.
\label{eq:isoamps}
\end{eqnarray}
In the CP violation case, $A_0$, $A_2$ and $A_2^+$ acquire imaginary parts and  $\varepsilon'$ can be written to first order in CP violation as
\begin{equation}
\varepsilon'\,  =\, 
- \frac{i}{\sqrt{2}} \: e^{i ( \chi_2 - \chi_0 )} \:\omega\;
%\frac{\mathrm{Re}\: A_{2}}{ \;\mathrm{Re}\: A_{0}} \,
\left[
\frac{\mathrm{Im}\: A_{0}}{ \mathrm{Re}\: A_{0}} \, - \,
\frac{\mathrm{Im}\: A_{2}}{ \mathrm{Re}\: A_{2}} \right] ,
\label{eq:cp1}
\end{equation}
showing that $\varepsilon'$ is suppressed by the ratio $\omega \equiv 
\mathrm{Re}\: A_2 / \mathrm{Re}\: A_0\approx 1/22$ and $\varepsilon'/\varepsilon$ is approximately real since $\chi_2 - \chi_0 - \pi/2\approx 0$ from Eqs.~(\ref{eq:isoamps}).
In addition, the CP-conserving amplitudes $\mathrm{Re} A_{I}$ are fixed to their experimental values in order to reduce the theoretical uncertainty. Consequently, a theoretical calculation is only needed for $\mathrm{Im} A_{I}$. 

In addition, Eq.~\eqn{eq:cp1} presents a delicated numerical balance between the two isospin contributions which makes the result very sensitive to the predictions of the CP-violating amplitudes. Hence, naive estimates of $\mathrm{Im} A_{I}$ give rise a strong cancellation between the two terms, leading to low values of $\varepsilon'/\varepsilon$ \cite{Buras:1993dy,Buras:1996dq,Bosch:1999wr,Buras:2000qz,Ciuchini:1995cd,Ciuchini:1992tj}.

Due to the ``$\Delta I =1/2$ rule'', the isospin breaking effects are very important in $\varepsilon'/\varepsilon$ \cite{Cirigliano:2003nn,Cirigliano:2003gt,Cirigliano:2009rr}. Including isospin violation, 
\be\label{eq:epsp_simp}
\mbox{}\hskip -.6cm
\mathrm{Re}\Bigl(\frac{\varepsilon'}{\varepsilon}\Bigr) \, = \, - \frac{\omega_+}{\sqrt{2}\, |\varepsilon|}  \, \left[
\frac{\mathrm{Im}\: A_{0}^{(0)} }{ \mathrm{Re}\: A_{0}^{(0)} }\,
\left( 1 - \Omega_{\rm eff} \right) - \frac{\mathrm{Im}\: A_{2}^{\rm emp}}{ \mathrm{Re}
  \: A_{2}^{(0)} } \right]  , \;\;
\ee
where $\mathrm{Im} A_2^{\rm emp}$ are the electromagnetic penguin operator contributions with $I=2$, $\omega_+ \equiv \mathrm{Re}\: A_{2}^{+}/\mathrm{Re} \: A_{0}$ and the superscript $(0)$ denotes the isospin limit. Furthermore, $\Omega_{\rm eff} = (6.0\pm 7.7) \cdot 10^{-2}$\cite{Cirigliano:2003nn,Cirigliano:2003gt} encodes all first order isospin-breaking corrections. An update of $\Omega_{\rm eff}$ is currently under way \cite{VincenzoEtAl}.

\section{$\chi$PT and determination of LECs}

At the electroweak scale, all flavour-changing transitions are described in terms of quarks and gauge bosons. The $K\to\pi\pi$ decay is a $\Delta S = 1$ transition in which due to the different mass scales ($M_\pi<M_K\ll M_W$), the gluonic corrections are amplified with large logarithms, $\log(M_W/m_c)\sim 4$. Using the Operator Product Expansion (OPE) and the Renormalization Group Equations (RGEs) in all the way down to scales $\mu<m_c$, one can sum up all these large logarithmic corrections and, finally, one gets an effective $\Delta S = 1$ Lagrangian defined in the three-flavour theory \cite{Buchalla:1995vs},
\be\label{eq:Leff}
\cL_{\mathrm{eff}}^{\Delta S=1}\, =\, - \frac{G_F}{\sqrt{2}}\,
 V_{ud}^{\phantom{*}}V^*_{us}\,  \sum_{i=1}^{10}
 C_i(\mu) \, Q_i (\mu)\, ,
\ee
which is a sum of local four fermion operators, $Q_i$, constructed with the light degrees of freedom and weighted by the Wilson coefficients $C_i(\mu)$ which are functions of the heavy masses and  CKM  parameters, $C_i(\mu)=z_i(\mu)+\tau\: y_i(\mu)$ with $\tau\equiv-\frac{V_{td}V_{ts}^*}{V_{ud}V_{us}^*}$ being the source of CP violation. The Wilson coefficients are known at NLO \cite{Buras:1991jm,Buras:1992tc,Buras:1992zv,Ciuchini:1993vr}. This means all corrections of $\cO(\alpha_s^n t^n)$ and $\cO(\alpha_s^{n+1} t^n)$, where $t\equiv\log{(M_1/M_2)}$ refers to the logarithm of any ratio of
heavy mass scales $M_{1,2}\geq\mu$. Some next-to-next-to-leading-order (NNLO) corrections are already known \cite{Buras:1999st,Gorbahn:2004my} and efforts towards a complete short-distance calculation at the NNLO are currently under way \cite{Cerda-Sevilla:2016yzo}.

$K^0\to\pi\pi$ is a very low energy process below the resonance region where perturbation theory no longer works. However, at this energy scale, one can use symmetry considerations to define another effective field theory in terms of Nambu-Goldstone bosons ($\pi$, $K$, $\eta$). 
Chiral Perturbation Theory ($\chi$PT)  describes the pseudoscalar octet dynamics through a perturbative expansion in powers of momenta and quark masses over the chiral symmetry breaking scale $\Lambda_\chi\sim 1$~GeV \cite{Weinberg:1978kz,Gasser:1983yg,Gasser:1984gg}. At lowest order, the most general nonleptonic electroweak Lagrangian contains three terms,
\begin{eqnarray}\label{eq:LchiPT}
\cL_{2}^{\Delta S=1}\:=\:G_8\:\cL_{8}\:+\:G_{27}\:\cL_{27}\:+\:G_8\:g_{\mathrm{ewk}}\:\cL_{\mathrm{ewk}}\, .
\end{eqnarray}
Then, $\cL_{2}^{\Delta S=1}$ determines the $K^0\to\pi\pi$ amplitudes at $\mathcal{O}(p^2)$ in terms of electroweak chiral couplings. In addition, $G_8$, $G_{27}$ and $G_8\, g_{\mathrm{ewk}}$ hide all the quantum information of heavy particles that are not dynamical at this regime and then they can not be fixed by the symmetries. A first-principle computation of these three LECs requires to perform a matching between the short-distance and effective  Lagrangians in Eqs.~(\ref{eq:Leff}) and (\ref{eq:LchiPT}). This can be easily done in the limit of an infinite number of quark colours, where the four-quark operators factorize into currents with well-known chiral realizations. Since the large-$N_C$ limit is only applied in the matching between the two effective field theories, the only missing contributions are $1/N_C$ corrections that are not enhanced by any large logarithms.

\section{Impact of {\boldmath $K\to\pi\pi$} amplitudes on {\boldmath $\varepsilon'/\varepsilon$}}

At LO in $\chi$PT, the phase shifts are predicted to be zero, because they are generated through loop diagrams with $\pi\pi$ absorptive cuts. 
The large value of the measured phase-shift difference in Eq.~\eqn{eq:isoamps} indicates a very large absorptive contribution. Analyticity relates the absorptive and dispersive parts of the one-loop diagrams, which implies that the dispersive correction is also very large.
A proper calculation of chiral loop corrections is then compulsory in order to obtain a reliable prediction for $\varepsilon'/\varepsilon$, since Eq.~(\ref{eq:cp1}) presents a strong cancellation between the two isospin contributions in simplified analyses. Naive estimates, which completely ignore the importance of these absortive cuts, obtain small SM values of $\epsilon'/\epsilon$ highlighting the importance of these contributions, since they are not able to predict a phase shift difference compatible with Eq.~(\ref{eq:isoamps}) \cite{Buras:2015xba,Buras:2016fys,Buras:2015yba}.

At the NLO in $\chi$PT, the $\cA_{\Delta I}$ amplitudes can be written in the form
\begin{eqnarray}
\mbox{}\hskip -.2cm
\mathcal{A}_{\Delta I} \, &\!\!\! =&\!\!\! -\,
G_8 F_\pi\,\Bigl\{ (M_K^2-M_\pi^2)\,\mathcal{A}_{\Delta I}^{(8)} -e^2 F_{\pi}^2 \, g_{\mathrm{ewk}}\,\mathcal{A}_{\Delta I}^{(g)}\Bigr\}
 -\, G_{27} F_\pi\, (M_K^2-M_\pi^2)\,\mathcal{A}_{\Delta I}^{(27)}\, ,
\label{amplitudegeneral}
\end{eqnarray}
where $\mathcal{A}_{\Delta I}^{(8)}$ and $\mathcal{A}_{\Delta I}^{(27)}$ represent the octet and 27-plet components, and $\mathcal{A}_{\Delta I}^{(g)}$ contains the electroweak penguin contributions. Moreover, these quantities can be further decomposed as
\be 
\cA_{\Delta I}^{(X)}\, =\, a_{\Delta I}^{(X)} \,\left[ 1 + \Delta_L\cA_{\Delta I}^{(X)} + \Delta_C\cA_{\Delta I}^{(X)}\right]\, ,
\ee
with $a_{\Delta I}^{(X)}$ the tree-level contributions, $\Delta_L\cA_{\Delta I}^{(X)}$ the one-loop chiral corrections and $\Delta_C\cA_{\Delta I}^{(X)}$ the NLO local corrections at $\mathcal{O}(p^4)$. 
The numerical values of  the different $\cA_{1/2}^{(X)}$ and $\cA_{3/2}^{(X)}$ components are displayed in tables~\ref{tab:Table12} and \ref{tab:Table32}, respectively.  

%%%%%%%%%%%%%%%%%%%%%%%% Tables O(p^4) Corrections %%%%%%%%%%%%%%%%%%%%%%%%
\begin{table}[hb] 
\renewcommand*{\arraystretch}{1.3}
\begin{center}
\begin{tabular}{|c|c|c|c|c|}\hline 
X & $\! a_{1/2}^{(X)}\! $ &
$\Delta_{L} \mathcal{A}_{1/2}^{(X)} $ &
$[\Delta_{C} \mathcal{A}_{1/2}^{(X)}]^+ $  & 
$[\Delta_{C} \mathcal{A}_{1/2}^{(X)}]^- $ \\
\hline   
\hline 
8  & $\sqrt{2}$   & $\! 0.27 +  0.47\,  i\! $   &
$\! \phantom{-}0.01 \pm 0.05\! $   &  $\! \phantom{-}0.02 \pm 0.05\! $  
\\
\hline 
g  & $ \! \frac{2 \sqrt{2}}{3}\! $  &  $\! 0.27 +  0.47\, i\! $ & 
$\! -0.19 \pm 0.01\! $  & $\! -0.19 \pm 0.01\! $ 
\\
\hline 
$\!\! 27\!\!$  &  $\frac{\sqrt{2}}{9}$   &  $\! 1.03 + 0.47\, i\! $    & 
$\! \phantom{-}0.01 \pm 0.63\! $ & $\! \phantom{-}0.01  \pm 0.63\! $ 
\\
\hline   
\end{tabular}
\caption{Numerical predictions for the $\mathcal{A}_{1/2}$ components. The local NLO correction to the CP-even ($[\Delta_{C} \mathcal{A}_{1/2}^{(X)}]^+$) and CP-odd ($[\Delta_{C} \mathcal{A}_{1/2}^{(X)}]^-$) amplitudes is only different in the octet case.} 
\label{tab:Table12} 
%\end{center} 
%\end{table}
\vskip .5cm
%%%%%%%%%%%%%%%%%%%%%
%
%\begin{table}[h!] 
%\renewcommand*{\arraystretch}{1.3}
%\begin{center}
\begin{tabular}{|c|c|c|c|}\hline 
X & $a_{3/2}^{(X)}$ &
$\Delta_{L} \mathcal{A}_{3/2}^{(X)} $ & $\Delta_{C} \mathcal{A}_{3/2}^{(X)} $
% $[\Delta_{C} \mathcal{A}_{3/2}^{(X)}]^+ $  & $[\Delta_{C} \mathcal{A}_{3/2}^{(X)}]^- $ 
\\
\hline   
\hline 
g  & $ \frac{2}{ 3}$  &  $-0.50 -  0.21\; i$  & $-0.19 \pm 0.19$      
\\
\hline 
27  &  $\frac{10}{9}$  &  $-0.04 - 0.21\; i$    & 
$\phantom{-}0.01 \pm 0.05$    
\\
\hline   
\end{tabular}
\caption{Numerical predictions for the $\mathcal{A}_{3/2}$ components. }  
\label{tab:Table32} 
\end{center} 
\end{table}
%%%%%%%%%%%%%%%%%%%%%%%%%%%%%%%%%%%%%%%%%%%%%%%%%%%%%%%%%%%%%%%%%%%%%%%%

The absorptive chiral corrections are large and positive for the $\Delta I=1/2$ amplitudes and much smaller and negative for $\Delta I=3/2$. Furthermore, they do not depend on the chiral renormalization scale $\nu_\chi$. Besides, table~\ref{tab:Table12} shows a huge dispersive one-loop correction to the $\cA^{(27)}_{1/2}$ amplitude. However, since Im$(g_{27})=0$, the 27-plet components do not contribute to the CP-odd amplitudes and, therefore, do not introduce any uncertainty in the final numerical value of Im$A_0$.

The relevant NLO loop corrections for $\varepsilon'/\varepsilon$ are $\Delta_L\cA_{1/2}^{(8)}$ and $\Delta_L\cA_{3/2}^{(g)}$. The first one generates a significant enhancement of Im$A_0$, $|1 +\Delta_L\cA_{1/2}^{(8)}|\approx 1.35$, while the second one produces a suppression in Im$A_2^{\mathrm{emp}}$, $|1 +\Delta_L\cA_{3/2}^{(g)}|\approx 0.54$. Consequently, the numerical cancellation between the $I=0$ and $I=2$ terms in Eq.~\eqn{eq:cp1} is completely destroyed by the chiral loop corrections.

Furthermore, tables~\ref{tab:Table12} and \ref{tab:Table32} show also the numerical predictions for the NLO local corrections $\Delta_C\cA_{\Delta I}^{(X)}$, which have been estimated in the large-$N_C$ limit. The most significant local corrections for $\varepsilon'/\varepsilon$ are $[\Delta_C\cA_{1/2}^{(8)}]^-$ and $\Delta_C\cA_{3/2}^{(g)}$; nevertheless, they are much smaller than the loop contributions. Further details can be found in \cite{Gisbert:2017vvj}.

\section{\boldmath The SM prediction for $\varepsilon'/\varepsilon$}

Taking into account all computed corrections in Eq.~\eqn{eq:epsp_simp}, our SM prediction for $\varepsilon'/\varepsilon$ is
\beqn\label{eq:finalRes}
\mbox{Re}\left(\varepsilon'/\varepsilon\right) =
\left(15\pm 2_{\mu}\pm 2_{m_s} \pm 2_{\Omega_\mathrm{eff}}\pm 6_{1/N_C}\right) \times 10^{-4}
  = 
\left(15\pm 7\right) \times 10^{-4}\, .
\eeqn
The first uncertainty has been estimated by varying the short-distance renormalization scale $\mu$ between $M_\rho$ and $m_c$. The second error shows the sensitivity to the strange quark mass, within its allowed range, while the third one displays the uncertainty from the isospin-breaking parameter $\Omega_{\mathrm{eff}}$. The last error is our dominant source of uncertainty and reflects our ignorance
about $1/N_C$-suppressed contributions that we have missed in the matching process.

\begin{figure}[t]\centering
\includegraphics[scale=.5]{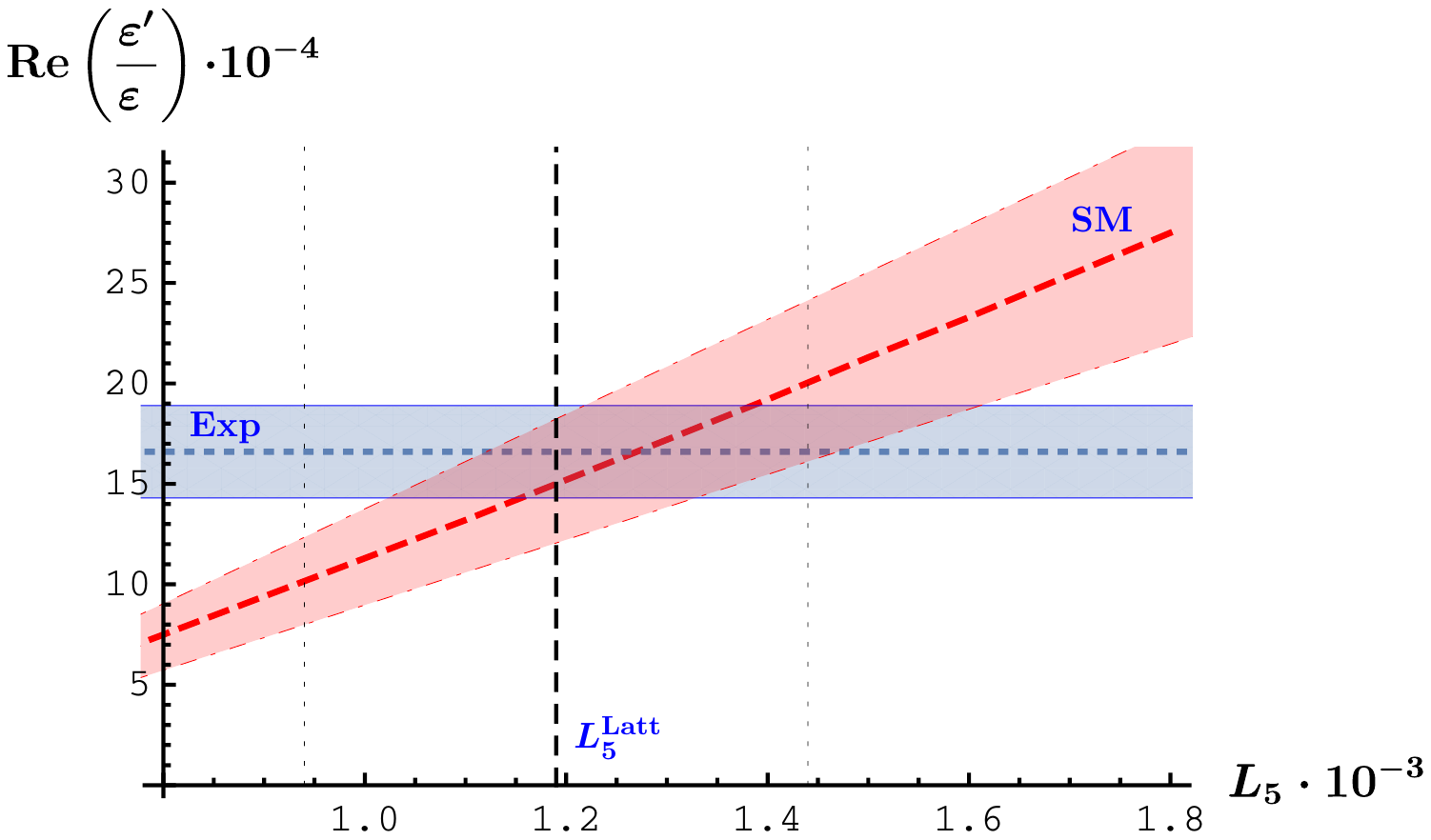}
\caption{SM prediction for $\varepsilon'/\varepsilon$ as function of $L_5$
(red dashed line) with $1\,\sigma$ errors (oblique band).
The horizontal blue band displays the experimentally measured value with $1\,\sigma$ error bars. The dashed vertical line shows the current lattice determination of $L_5^r(M_\rho)$.}
\label{fig:epsp}
\end{figure}

In figure~\ref{fig:epsp}, we plot the prediction for $\varepsilon'/\varepsilon$ as function of the $\chi$PT coupling $L_5$, which clearly shows a strong dependence on this parameter. The experimental $1\,\sigma$ range is indicated by the horizontal band, while the dashed vertical lines display the current lattice 
determination of $L_5^r(M_\rho)$. The measured value of $\varepsilon'/\varepsilon$ is nicely reproduced with the preferred lattice inputs. The current error on $L_5$ is the largest parametric contribution to the $1/N_C$ uncertainty in (\ref{eq:finalRes}).

%%%%%%%%%%%%%%%%%%%%%%%%% Figure %%%%%%%%%%%%%%%%%%%%%%%%%%%%%%%%%%%%%%%%
%%%%%%%%%%%%%%%%%%%%%%%%%%%%%%%%%%%%%%%%%%%%%%%%%%%%%%%%%%%%%%%%%%%%%%%%%

Our SM prediction for $\varepsilon'/\varepsilon$ is in perfect agreement with the measured experimental value. We have shown the important role of FSI in $K^0\to\pi\pi$. When $\pi\pi$ re-scattering corrections are taken into account, the numerical cancellation between the $Q_6$ and $Q_8$ terms in Eq.~\eqn{eq:cp1} is completely destroyed because of the positive enhancement of the $Q_6$ amplitude and the negative suppression of the $Q_8$ contribution. Once these important corrections are included, the contributions from other four-quark operators to $\mathrm{Im} A_0^{(0)}$ and $\mathrm{Im} A_2^{\mathrm{emp}}$ become numerically less relevant, since  the cancellation is no longer operative.

The claims \cite{Buras:2015xba,Buras:2016fys,Buras:2015yba} of a flavour anomaly in $\varepsilon'/\varepsilon$ are either based on the recent lattice simulation that fails to reproduce the correct phase shifts or they  originate in naive approximations that overlook the important role of pion chiral loops. These incorrect estimates are using simplified ansatzs for the $K\to\pi\pi$ amplitudes, without any absorptive contributions, in complete disagreement with the strong experimental evidence of a very large phase shift difference.

Our SM prediction of $\varepsilon'/\varepsilon$ agrees well with the measured value and provides a qualitative confirmation of the SM mechanism of CP violation. Although the theoretical error is still large, improvements can be achieved  in the next years via a combination of analytical calculations, numerical simulations and data analyses \cite{Gisbert:2017vvj}. 

\section*{Acknowledgements}

I want to thank the organizers for their effort to make this conference such a successful event. I also thank Vincenzo Cirigliano, Gerhard Ecker, Elvira Gámiz, Helmut Neufeld, Toni Pich, Jorge Portolés and Antonio Rodríguez Sánchez  for useful discussions. This work has been supported in part by the Spanish State Research Agency and ERDF funds from the EU Commission [Grants FPA2017-84445-P and FPA2014-53631-C2-1-P], by Generalitat Valenciana [Grant Prometeo/2017/053], by the Spanish Centro de Excelencia Severo Ochoa Programme [Grant SEV-2014-0398] and by a FPI doctoral contract [BES-2015-073138], funded by the Spanish
State Research Agency.

\end{document}